\documentclass[aps,prl,twocolumn,groupedaddress,superscriptaddress,showpacs]{revtex4}

\usepackage{graphicx}% Include figure files
\usepackage{dcolumn}% Align table columns on decimal point
\usepackage{bm}% bold math
\usepackage[usenames]{color}
\usepackage{ulem}

\begin{document}

\title{New interpretation of the origin of 2DEG states at the surface of layered topological insulators}

\author{S.V. Eremeev}
\affiliation{Tomsk State University, 634050, Tomsk, Russia\\}
\affiliation{Institute of Strength Physics and Materials Science,
634021, Tomsk, Russia\\}

\author{T.V. Menshchikova}
 \affiliation{Tomsk State University, 634050, Tomsk, Russia\\}

\author{M.G. Vergniory}
 \affiliation{Donostia International Physics Center (DIPC),
             20018 San Sebasti\'an/Donostia, Basque Country,
             Spain\\}

\author{E.V. Chulkov}
\affiliation{Donostia International Physics Center (DIPC),
             20018 San Sebasti\'an/Donostia, Basque Country,
             Spain\\}
\affiliation{Departamento de F\'{\i}sica de Materiales UPV/EHU,
Centro de F\'{\i}sica de Materiales CFM - MPC and Centro Mixto
CSIC-UPV/EHU, 20080 San Sebasti\'an/Donostia, Basque Country, Spain
\\}

\date{\today}

\begin{abstract}
On the basis of relativistic ab-initio calculations we show that
the driving mechanism of simultaneous emergence of parabolic and
M-shaped 2D electron gas (2DEG) bands at the surface of layered
topological insulators as well as Rashba-splitting of the former
states is an expansion of van der Waals (vdW) spacings caused by
intercalation of metal atoms or residual gases. The expansion of
vdW spacings and emergence of the 2DEG states localized in the
(sub)surface region are also accompanied by a relocation of the
topological surface state to the lower quintuple layers, that can
explain the absence of interband scattering found experimentally.
\end{abstract}


\maketitle

The recently discovered three-dimensional topological insulators
belong to a class of insulators in which the bulk gap is inverted
due to the strong spin-orbit interaction
\cite{Fu_PRL_07,Fu_PRB_07,Qi_PRB_08,Zhang_NatPhys_09}. A direct
consequence of such bulk band structure arises at the surface: the
spin-polarized topologically protected massless metallic states,
forming a Dirac cone. These surface states (SS) exhibit many
interesting properties resulting from the fact that the spin of
electron is locked perpendicular to its momentum, thus forming a SS
spin structure that protects electrons from backscattering under the
influence of weak perturbations. This makes topological insulators
potentially promising materials for creation of new quantum devices.

By now few families of topological insulators (TI) have been
discovered \cite{Shitade,Chadov,Lin_MN2010,EremeevJETPLettTh,Lin_PRL2010,%
Kuroda_PRL2010,EremeevJETPL_PBT,Xia_NatPhys2009,ChenScience09,ZhangPRL09,KurodaPRL10},
of which the binary layered compounds Bi$_2$Te$_3$, Bi$_2$Se$_3$,
and Sb$_2$Te$_3$ are the most studied both experimentally
\cite{Xia_NatPhys2009,ChenScience09,ZhangPRL09,KurodaPRL10} and
theoretically
\cite{Zhang_NatPhys_09,EremeevJETPL_10,ZhangNatPhys10,SongPRL10,ZhangNJP10,Yazyev_PRL2010}.
The latter systems have tetradymite-like layered structure with
ionic-covalent bonded quintuple layer (QL) slabs, which are linked
by weak van der Waals forces. Such a layered structure predetermines
the formation of the surface by cleavage on the van der Waals (vdW)
spacing that doesn't result in the formation of dangling bonds, and
thus only the Dirac states reside in the bulk energy gap.

However, recently in several studies it has been demonstrated by
using angle-resolved photoemission spectroscopy (ARPES) that beside
the Dirac cone the 2DEG states arise at the surface of Bi$_2$Se$_3$
after a few hours of exposition in vacuum \cite{Bianchi}, upon
deposition of various magnetic \cite{Wray_NatPhys11,Valla} and
non-magnetic atoms \cite{Valla,Wray_arXiv,Zhu_arXiv}, and molecules
as well \cite{Benia_arXive,Wray_arXiv}. These states form a
parabolic band (PB) in the energy gap just below the conduction band
and M-shaped band in the local gap of bulk-projected valence band.
The former bands show an appreciable Rashba spin splitting
\cite{King_Arxiv,Wray_NatPhys11,Valla,Benia_arXive,Zhu_arXiv}.
Moreover, for several adsorbates \cite{Valla,Zhu_arXiv} at the
saturation deposition time a second and even third pair of
spin-split parabolic states emerges below the conduction band
minimum. In the most of these papers the emergence of PB states was
ascribed to a confinement of the conduction band states in a quantum
well formed by band bending potential produced by adsorption of
metallic atoms or residual gases, although in
Ref.~\cite{Benia_arXive} it was pointed out that the potential
gradient from band bending cannot alone be responsible for their
Rashba-splitting. Indeed, a model calculation exploiting band
bending approach based on the coupled solution of the Poisson and
Schr\"{o}dinger equations yielded the Rashba spin-orbit coupling
parameter $\alpha_R$ which is significantly smaller than that
extracted from the ARPES data \cite{King_Arxiv}. Moreover this
approach doesn't reproduce the M-shaped band \cite{Bianchi}. The
most surprising experimental finding is the absence of interband
scattering of the Dirac state electrons in the presence of the PB
spin-split states after the deposition of any kind of metal atoms
\cite{Valla}. A study of a naturally aged surface of Bi$_2$Se$_3$
also reveals that scattering rate in the topologically protected
state is unaffected by the potential created by adsorbed atoms or
molecules \cite{Park_PRB10}.

In this Letter we propose an alternative explanation of the
emergence of the 2DEG states based on a well known fact: the
interlayer gaps (vdW spacings) in the layered compounds can serve as
natural containers for impurities in synthesis processes and for
intercalated atoms. Various atoms have been  intercalated
in different layered materials, inducing expansion of vdW
spacings \cite{Dresselhaus,Friend,ILM}. One would expect that due to
a weak binding between QLs even a relatively small concentration of
contaminants in the vdW gap can produce its sizeable expansion. The
magnitude of this expansion depends on the impurity size and on
formation of impurity atom clusters within the vdW gap. Mechanisms
of diffision of adsorbed atoms into the vdW gap are not known in
detail yet, however, at least two kinds of the diffusion process can
be anticipated. First, the Fermi level of naturally grown crystals
of the tetradymite-like TI is usually found to be located in the
bulk conduction band due to vacancy defects
\cite{Xia_NatPhys2009,KurodaPRL10} and, thus, apparently there is a
sufficiently high vacancy concentration to ensure the vacancy
mediated penetration of impurities through a QL into the vdW gap.
The second process can be associated with the sliding of impurity
atoms into the vdW gap from step edges.

Here we show on the basis of ab-initio calculations using the VASP
code \cite{VASP1,VASP2} that the driving mechanism of the
simultaneous emergence of the parabolic and M-shaped bands in
Bi$_2$Se$_3$, Bi$_2$Te$_3$, and Sb$_2$Te$_3$, is a widening of the
outermost vdW spacing. This scenario also explains the Rashba-type
splitting for the PB. We find that besides the emergence of PB
states, which are localized in the detached QL the expansion of the
vdW spacing results in a relocation of the Dirac state to the lower
QL that makes topological and PB states separated in space. This
fact explains the observable absence of interband scattering
\cite{Valla,Park_PRB10}. We also show that the expansion of various
vdW gaps produces multiple 2DEG states.

\begin{figure}
\begin{center}
\includegraphics[width=\columnwidth]{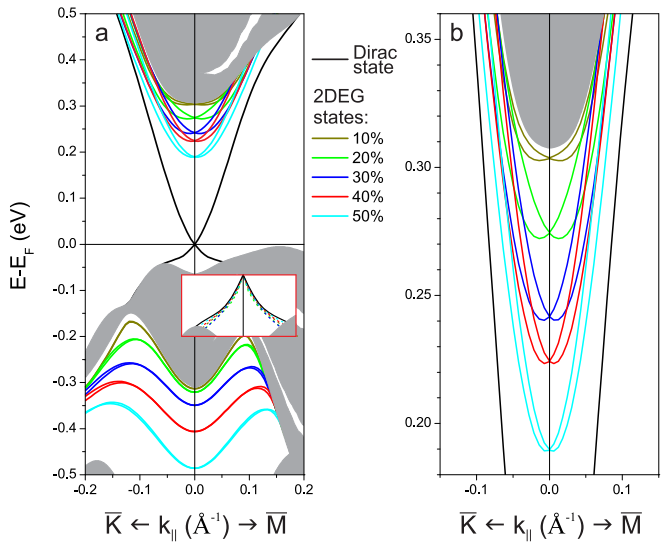}
\caption{(Color online) (a) Surface band structure of Bi$_2$Se$_3$
in the vicinity of $\bar \Gamma$ with the vdW gap expanded from 10\%
to 50\% (insert: variation of the lower part of the Dirac cone with
vdW gap expansion); (b) a magnified view of Rashba-split parabolic
bands.}
 \label{fig:Bi2Se3bnd}
\end{center}
\end{figure}

\begin{table}
\caption{The energy with respect to $E_{\rm F}$, the fitted effective
mass, and the Rashba coupling parameter $\alpha_{R}$ for the parabolic
SS as a function of the vdW gap expansion. }
\begin{ruledtabular}
\begin{tabular}{ccccccc}
  &expansion & $E_0$ & \multicolumn{2}{c}{$m^*$} & \multicolumn{2}{c}{$\alpha_{R}$ (eV$\cdot$\AA)}  \\
  &(\%)& (eV) &$\bar\Gamma$-$\bar{\rm K}$&$\bar\Gamma$-$\bar{\rm M}$&$\bar\Gamma$-$\bar{\rm K}$&$\bar\Gamma$-$\bar{\rm M}$\\
 \hline
Bi$_2$Se$_3$ & 10 &0.30&0.38&0.46&0.16&0.14\\
             & 20 &0.27&0.27&0.28&0.25&0.24\\
             & 30 &0.24&0.20&0.20&0.33&0.32\\
             & 40 &0.22&0.18&0.18&0.29&0.29\\
             & 50 &0.19&0.17&0.17&0.24&0.24\\
 \hline
Bi$_2$Te$_3$ & 10 &0.15&0.31&0.37&0.29&0.28\\
             & 20 &0.04&0.17&0.18&0.40&0.40\\
             & 30 &0.01&0.14&0.13&0.74&0.71\\
             & 40 &0.00&0.12&0.12&0.59&0.57\\
             & 50&-0.02&0.12&0.13&0.45&0.45\\
 \hline
Sb$_2$Te$_3$ & 10 &0.09&0.24&0.28&0.30&0.30\\
             & 20 &0.08&0.17&0.18&0.35&0.36\\
             & 30 &0.05&0.12&0.13&0.26&0.26\\
             & 40 &0.04&0.12&0.12&0.15&0.15\\
             & 50 &0.02&0.12&0.12&0.10&0.10\\
\end{tabular}
\end{ruledtabular}
\label{tab:tab1}
\end{table}

We start with consideration of the Bi$_2$Se$_3$ surface. To simulate
the effect of the adsorbate deposition time as well as the impurity
atom size we perform a calculation of the Bi$_2$Se$_3$ surface with
expansion of the outermost vdW spacing by 10\% - 50\%. As one can
see (Fig.~\ref{fig:Bi2Se3bnd}), the detachment of the outermost QL
leads to the simultaneous emergence of both the Rashba-split band
below the bottom of the bulk conduction band and the M-shaped band
in the valence bulk-projected gap. The energy of the M-shaped and
parabolic spin-split bands as well as the magnitude of the Rashba
splitting parameter strongly depend on the vdW spacing expansion. At
10\% expansion the first pair of spin-split bands as well as the
M-shaped state emerges just below the bulk projected states. Upon
increasing the vdW expansion these bands shift gradually down. This
behavior reflects the rise of the 2DEG bands as a function of the
deposition time found in recent experimental studies
\cite{Valla,Benia_arXive}. Up to 30\% expansion this also leads to
the decrease of the effective mass $m^*$ of PBs and to the increase
of the Rashba coupling parameter $\alpha_{R}$ (see
Table~\ref{tab:tab1}). At the same time $m^*$ demonstrates an
apparent $\bar\Gamma$-$\bar{\rm K} / \bar\Gamma$-$\bar{\rm M}$
anisotropy at 10\% expansion. At higher expansions this anisotropy
is significantly reduced. The small ($\approx 0.01$) $m^*$
anisotropy was also obtained in Ref.~\cite{King_Arxiv}(SI).
Different fitting procedures for the ARPES measured band in
Ref.~\cite{King_Arxiv}(SI) gave the Rashba splitting parameter
varying from 0.36 to 1.35 eV$\cdot$\AA\ in aged Bi$_2$Se$_3$
surface. In the case of potassium deposition on Bi$_2$Se$_3$ a
maximal $\alpha_R=0.79\pm0.03$ eV$\cdot$\AA\ was observed after
2.5-3 minute K evaporation \cite{Zhu_arXiv}. Our model gives a lower
limit for the former case at moderate expansions and underestimates
the latter maximal $\alpha_R$ while the band bending model
\cite{King_Arxiv} gives 0.1 eV$\cdot$\AA\ for the PB splitting.

\begin{figure}
\begin{center}
\includegraphics[width=\columnwidth]{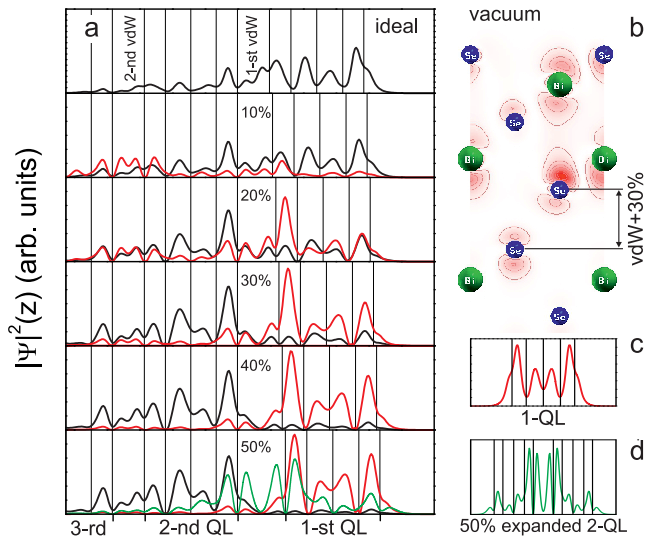}
\caption{(Color online) (a), Spatial localization of the topological
(black) and parabolic (red) surface states as a function of the vdW
gap expansion; green line represents localization of the M-shaped
band for 50\% vdW gap expansion; (b), charge density distribution of
the PB state for 30\% expansion; (c) and (d), $|\Psi(z)|^2$ of the
lowest unoccupied band for a free-standing 1-QL and of the upper
occupied band for a free-standing 2-QL with 50\% vdW gap expanded,
respectively.}
 \label{fig:Bi2Se3chg}
\end{center}
\end{figure}

% $|\Psi|^2(z)$

The nature of spin-split PBs can be deduced from analysis of their
spacial localization. At 10\% expansion the emerged parabolic state,
which splits off from the conduction bulk band continuum, keeps yet
a bulk-like delocalized character (Fig.~\ref{fig:Bi2Se3chg}(a)).
This explains the obtained $\bar\Gamma$-$\bar{\rm K} /
\bar\Gamma$-$\bar{\rm M}$ anisotropy of the parabolic band at small
expansions: it just reflects a similar anisotropy of the bulk
states. Upon increasing the vdW spacing this state acquires more
localized character and at 30-50\% it almost completely lies in the
detached QL.
%The PB state, characterized by four peaks localized in
%the outermost QL, Fig.~\ref{fig:Bi2Se3chg}(a), is composed of the Bi
%and Se orbitals with the sizeable Se $p_z$ contribution to the
%vdW-side peak (Fig.~\ref{fig:Bi2Se3chg}(b)).
The PB state, characterized by four peaks localized in the outermost
QL, Fig.~\ref{fig:Bi2Se3chg}(a), reflects the symmetry of the energy
gap edges at $\bar\Gamma$: it is composed of the Bi $p_z$ orbitals
(lower edge of the gap) with the sizeable Se $p_z$ contribution
(upper edge of the gap) to the vdW-side peak
(Fig.~\ref{fig:Bi2Se3chg}(b)).
 With the further increase of the
outermost vdW spacing the localization of the parabolic state
approaches the distribution $|\Psi(z)|^2$ of the lowest unoccupied
band in a free-standing QL (Fig.~\ref{fig:Bi2Se3chg}(c)). The latter
band is not spin-orbit split since a free-standing QL possess
inversion symmetry. This effect accounts for the reduction of
$\alpha_{R}$ in the PB states with the increase of the vdW spacing
beyond 30\% (Table~\ref{tab:tab1}). Note here that a large
concentration of intercalated atoms in the vdW gap can affect the
vdW-side Se orbitals and hence lead to change of some features of
the spin-split band, for instance increase the Rashba splitting.

In the case of Bi$_2$Te$_3$ and Sb$_2$Te$_3$ we find the same trend
in the development of the Rashba-split PB (Table~\ref{tab:tab1}).
However, the heavier Te provides two times bigger $\alpha_{R}$ in
Bi$_2$Te$_3$ with respect to that in Bi$_2$Se$_3$ while the lighter
Sb, in turn, gives a smaller Rashba coupling parameter in
Sb$_2$Te$_3$.

An intriguing finding is that the development of PB states is
accompanied by shifting the topological state deep into the crystal,
so that at 30-50\% expansion the Dirac state is mostly located in
the second QL, beneath the detached QL. In this case wave functions
of the topological and parabolic states only slightly overlap. The
latter explains the experimentally observed absence of interband
scattering \cite{Valla,Park_PRB10}. It should be stressed that the
dispersion of the upper part of the Dirac cone as well as the
position of the Dirac point remain unchanged under this relocation
while the dispersion of the lower part of the Dirac cone is slightly
modified (Fig.~\ref{fig:Bi2Se3bnd}(a), insert) in agreement with
ARPES data \cite{Wray_arXiv}.

The M-shaped band emerges in the local gap of the bulk-projected
valence band splitting off from the upper edge of the gap. The edge
is formed by the bulk Se $p_z$ states only which are not spin-orbit
split at $\bar\Gamma$ and around. The  M-shaped SS  being split off
from the Se $p_z$ bands does not acquire other symmetry and thus
does not show spin-orbit splitting at small momenta. Only being
close to the bulk continuum states (at large momenta) it shows weak
spin-orbit splitting. With the increase of the vdW gap it becomes
more localized. Charge density of this state as shown for 50\%
expansion case in Fig.~\ref{fig:Bi2Se3chg}(a) is mostly situated in
the expanded vdW gap and around. Comparing this $|\Psi(z)|^2$ with
that of the the upper occupied state in free-standing 2-QL with 50\%
vdW gap expanded, Fig.~\ref{fig:Bi2Se3chg}(d), one can note that
these two states are of the same origin: the vdW spacing expansion.
%Contrary to the PB these SS's are formed
%predominantly by $p_z$ orbitals of Se atoms bordering the expanded
%vdW gap that provides its spin unsplit character.

\begin{figure}
\begin{center}
\includegraphics[width=\columnwidth]{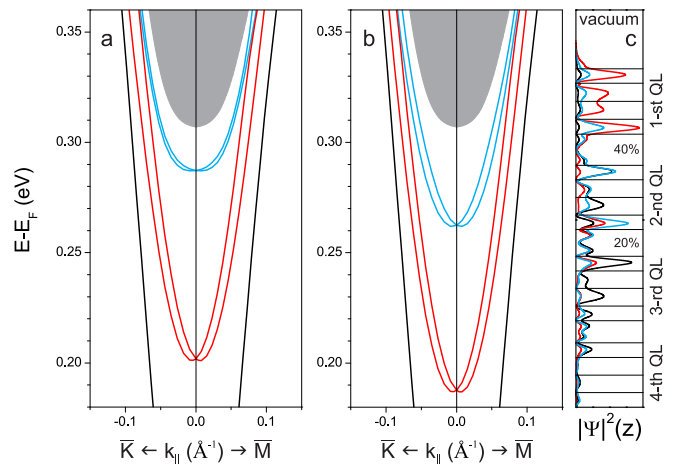}
\caption{(Color online) Surface band structure of Bi$_2$Se$_3$ with
the first and second vdW gaps expanded by 40\% and 10\%, respectively
(a); by 40\% and 20\%, respectively (b); spatial localization of
the topological state (black line) as well as for the  lower (red) and
upper (blue) Rashba-split SSs for the (b) case (c). }
 \label{fig:Bi2Se3_2QL}
\end{center}
\end{figure}

\begin{figure}
\begin{center}
\includegraphics[width=\columnwidth]{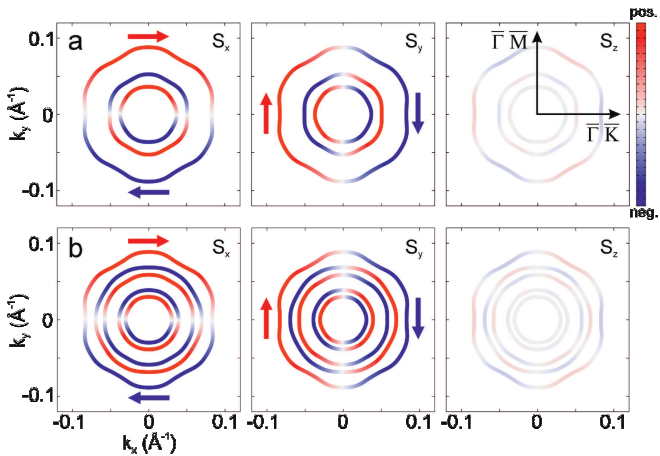}
\caption{(Color online) Spin structure of the Dirac state and
Rashba-split states of Bi$_2$Se$_3$, as given by spin projections
$S_x$, $S_y$, and $S_z$ at energy of 280 meV for the detached 1QL
(30\%) case (a) and for the 2QL (40\% + 20\%) case (b). }
 \label{fig:Bi2Se3spin}
\end{center}
\end{figure}

Now we consider the effect of simultaneous expansion of the first
and second vdW spacings. The expansion of the latter can be caused
by further diffusion in the sample of a smaller amount of atoms
deposited on the surface. In Fig.~\ref{fig:Bi2Se3_2QL}(a) we show
the band spectrum of Bi$_2$Se$_3$ with 40\% and 10\% expanded 1-st
and 2-nd vdW spacings, respectively. This film geometry gives rise
to the second parabolic state below the conduction band. This state
shows very small spin splitting. The emergence of the second PB
leads to a decrease of energy of the first band by 20 meV as well as
a reduction of its $\alpha_{R}$ to 0.27 eV$\cdot$\AA\ compared to
the case of the single vdW spacing expansion by 40\%. Further
increase of the 2-nd vdW gap leads to a subsequent downshift of both
PBs (Fig.~\ref{fig:Bi2Se3_2QL}(b)). While $\alpha_{R}$ for the upper
band in this case is 0.16 (0.15) eV$\cdot$\AA\ in
$\bar\Gamma$-$\bar{\rm K}$ ($\bar\Gamma$-$\bar{\rm M}$) directions,
respectively, the spin splitting of the lower state is reduced to
0.25 eV$\cdot$\AA. Localizations of the lower and upper bands in
general resemble those for a single vdW expansion by 40\% and 20\%,
respectively except their mutual overlap
(Fig.~\ref{fig:Bi2Se3_2QL}(c)). The overlap of the topological SS
with that localized in the 2-nd QL is similar to that in the single
20\% case while the Dirac state is absent in the outermost QL as in
the case of the single 40\% detachment. Further increase of the 2-nd
vdW spacing leads to bigger downshift of both PBs, to increase of
their overlap and complete relocation of the Dirac state to the
third QL. The expansion of the second vdW spacing also results in
the development of the second M-shaped state in the valence band gap
(not shown). Subsequently expansion of the third vdW gap, that
simulate a deeper diffusion of deposited atoms, leads to
simultaneous emergence of the third pair of parabolic Rashba-split
and M-shaped bands (not shown).

Note that the sizeable overlap between the topological SS and 2DEG
state exists barely at the beginning of the formation of each new PB
and hence the interband scattering is possible in a narrow energy
interval just below the conduction band only. In fact this
scattering will be also limited by spin conservation rule due to
spin polarized nature of both topological and 2DEG Rashba-split
states. The Dirac state has clockwise spin helicity while the
parabolic states demonstrate typical Rashba-type counter-clockwise
and clockwise helicities for outer and inner branches, respectively,
with a very small $S_z$ spin component for both types of SS's
(Fig.~\ref{fig:Bi2Se3spin}(a)). This spin structure allows
``$\mathbf k_1$ to $\mathbf k_2$" scattering between the topological
state and the inner branch of the PB and ``$\mathbf k_1$ to
$-\mathbf k_2$" transitions between the Dirac cone state and the
outer branch of the PB. However, the efficiency of these transitions
should be significantly reduced because of small overlap between the
topological SS and PB state for well developed PBs. In the case of
multiple detached vdW gaps new channels for scattering of Dirac
electrons (Fig.~\ref{fig:Bi2Se3spin}(b)) arise, nevertheless, only
PBs that have sizeable overlap with the topological SS can
contribute to considerable interband scattering.

To summarize, on the base of the proposed scenario of impurity
intercalation in the vdW gaps of the layered TIs and ab-initio
calculation results we have shown that the driving mechanism of
experimentally observed simultaneous emergence of parabolic and
M-shaped bands as well as Rashba-splitting of former bands is
expansion of vdW spacings. These bands are 2DEG states localized in
the narrow (sub)surface region. We have found that besides the
development of 2DEG SS's the expansion of the vdW gap also provides
relocation of the topological state to the lower QL. This fact
explains the observed absence of interband scattering between the
topological SS and Rashba-split PB states. We also show that
expansion of the low-lying vdW gaps produces multiple 2DEG states.
The proposed mechanism is expected to be also valid for other
layered TIs.

We acknowledge partial support by the University of the Basque
Country (project GV-UPV/EHU, grant IT-366-07) and Ministerio de
Ciencia e Inovaci{\'o}n (grant FIS2010-19609-C02-00). We thank
Ph.~Hofmann, A.~Kimura, Yu.M. Koroteev, and E.E.~Krasovskii for
enjoyable discussions.

\end{document}